\documentclass{mem}
\ifx\pdfoutput\undefined
\usepackage[dvips]{graphicx}
\else
\usepackage{graphicx}
\fi
\usepackage{epstopdf}

\usepackage{natbib}\usepackage{txfonts}\usepackage{balance}
\usepackage{graphicx}
\usepackage[a4paper]{hyperref}
\idline{00}{00}
\begin{document}
\DeclareGraphicsExtensions{.pdf,.gif,.jpg}
\def\teff{$T\rm_{eff }$}
\def\kms{$\mathrm {km s}^{-1}$}

\title{
VST processing facility: first astronomical applications
}

   \subtitle{}

\author{
A. \,Grado,\inst{1}\,
M. \,Capaccioli,  \inst{2, }\inst{3}\,
L. \,Limatola\inst{1}
\and F. \,Getman\inst{1}
          }

  \offprints{A. Grado}

\institute{
Istituto Nazionale di Astrofisica --
Osservatorio Astronomico di Capodimonte,
Via Moiariello, 16, 80131, Napoli, Italy
\email{agrado@na.astro.it}
\and
Universit\'a Degli Studi Federico II --
Complesso Monte Sant'Angelo
Napoli, Italy
\and
MECENAS, Universit\'a di Napoli Federico II and Universit\'a di Bari, Italy
}
\authorrunning{Grado }

\titlerunning{VST processing facility}

\abstract{VST--Tube is a new software package designed to process
optical astronomical images. It is an automated pipeline to go
from the raw exposures to fully calibrated co-added images, and to
extract catalogs with aperture and PSF photometry. A set of tools
allow the data administration and the quality check of the
intermediate and final products. VST-Tube comes with a Graphical
User Interface to facilitate the interaction between data and
user. We outline here the VST--Tube architecture and show some
applications enlightening some of the characteristics of the
pipeline.  } \maketitle{}

\section{Introduction}
Several pipelines have been developed in the last years to process astronomical
images. Most of them are tailored to specific instruments; others
are more flexible and can be applied to a variety of instruments.
A non--exhaustive list includes the SDSS pipeline \citep{lup01}, CASU
\citep{Irwin}, Astrowise \citep{valentijn}, EIS  \citep{hoo01},
TERAPIX \citep{swarp}, Theli \citep{sch07}, MegaPipe
\citep{gwy09}, NEWFIRM \citep{swa09}, QUEST \citep{and08},
Pan-STARRS  \citep{mag05}, SUBARU \citep{ogasawara}, MACHO
\citep{axe98}.

Here we present a new software package called VST--Tube, developed
to process raw astronomical images (typically in the optical
bands) and to make them suitable for scientific exploitation. It
has been conceived (but is not restricted) to process mosaicised
images such as those that will be produced by OmegaCam, the
imaging camera of the VLT Survey Telescope (VST) \citep{capacciolimessenger}. 
The VST--Tube pipeline allows automatic processing (from the raw
frames to a stacked image, astrometrically and photometrically
calibrated) of a set of exposures distributed in one or more
observing nights, limiting the user intervention to the pipeline
configuration and to the data preparation stages. A series of LOG
files keep track of the various input parameters and of the
configuration, so to allow reproducible results.

VST--Tube can run on a normal laptop as well as in a distributed 
processing environment such as a Beowulf cluster (more effective 
for mosaicised exposures). It comes with a graphical user interface
(VST-GUI) to simplify the usage and the access to intermediate and
final results, and to all the quality control checks done during
the processing, offering an intuitive approach through the menu
bar at the window top. Some additional tools are already available
to analyze the raw data and the pipeline results. In order to make
the approach even simpler, a series of balloons and messages pops
appear at pointing the mouse on almost the totality of the widget
windows.

In this paper we describe the software architecture and show three
examples of successfully processed images which enlighten the
characteristics of VST-Tube.

\section{VST--Tube pipeline: general description}
VST--Tube has been designed to process homogenous sets of
astronomical exposures, that is sets of images taken with the same
instrument and instrumental setup, but under different night sky
conditions. The input data are the science and calibration
exposures acquired with a single or a multi-chip optical camera,
the instrument and the pipeline configuration
files, and a certain number of options to be chosen at the
beginning in order to define the processing strategy. The output
are co-added images where the instrumental effects have been
removed. The latter include overscan and bias correction, flat-field
correction, CCDs mosaic gain harmonization, fringing pattern
removal (where applicable), illumination correction, relative and
absolute astrometric and photometric calibrations. 
Weight and flag maps, including cosmic ray traces, are also produced. 
Eventually, a catalog of sources
with aperture and PSF photometry can be automatically extracted.

The VST--Tube pipeline is configured through a Graphical
User Interface (GUI).
Written in PYTHON, it uses C libraries for Fits files manipulation
and external packages for calibration and quality assessment. Just
the GUI and the pipeline Python codes, excluding external packages
and libraries, consist of about 60,000 lines of code. 

VST--Tube has been conceived as a general purpose pipeline, {\it
i.e.\/} applicable to various telescopes and to rather different
science cases. In fact, the characteristics of VST--Tube are not only
speed and reliability but also versatility in either treating
different problems or in easily integrating user provided special
subroutines. It has been designed to satisfy the general
requirements of producing catalogs of astronomical sources with
point--source broad--band photometry with an absolute precision of 
$\sim 3 \%$ and a relative precision of $\sim 1 \%$ for repeated
observations. The absolute astrometric accuracy is limited by that
of the reference catalog (typically $\sim 0.3\  {\rm arcsec}$), while
the relative accuracy for repeated observations is $\sim 0.03 \ 
{\rm arcsec}$. It must be clear that the level of accuracy in the final
calibration depends on the characteristics of the science and 
calibration files.

The following is a short account of how VST--Tube works. Details
on the structure and on the recipes will be reported in
forthcoming papers.

\subsection{VST--Tube in action}
VST--Tube pipeline allows the user to create his own reduction strategy. 
Combining a number of parameters and reduction settings, the pipeline can 
be configured to meet the user needs and scopes.
In a typical session, the human intervention, all made through the GUI, 
consists in loading the instrument and the pipeline configuration files, 
selecting all the images of a given target that we want to reduce 
(in a given filter but not necessarily in just one observing night), 
choosing the method to produce the master calibration files
(there are several options depending on the set of available
data), selecting the kind of output we want to obtain (for
example, the creation of a single final mosaic or of a mosaic for each
exposure), and then switching on the reduction.  

The data input of the pipeline, properly prepared (see later), are indicated 
as the Object. The name Object is both used to designate the target and the tree 
directories created in the preparation phase, logging and archiving the 
intermediate and final products of the reduction process. 

Let us assume that the images we want to process come from a mosaicised instrument. 
In the zero--th step the data, in Multi-Extention Fits (MEF) format, are prepared 
to be processed. A Fits header analyzer, according to the instrument configuration 
parameters, catalogues the files and puts them in lists that identify the type of 
image. For each night and filter, the raw data will be classified in Bias, 
Flat, Dome, Dark, Standard, Science. A directory tree structure is created 
which contains the images, already transformed in Single-Extention-Fits (SEF), 
and all the intermediate and final pipeline products. A series of quality 
control checks are carried out on the images in order to reject those which do
not satisfy a priori requirements.
At this stage malformed Fits header keywords, where applicable, are fixed, and the World Coordinate 
System keyword values corrected in case of an inaccurate telescope pointing 
model. Quite often it happens that a header keyword, even for the same 
instrument, keeps changing during the instrument life. It is possible to 
define aliases in the instrument configuration file for some of the main 
keywords in order to identify univocally that keyword. Other simple checks, 
such as exposure time and median pixel value level, are checked against the 
allowed range defined in the instrument configuration file.
After this preparatory phase the images can be processed. All of the pipeline 
products belonging to one run are identified through a RUN-ID, which is the 
time when the processing started.
This string is added to the names of all the pipeline products to easily 
identify them even without the help of the GUI.

While drawing the main characteristics of our pipeline, we decided to use 
public domain software whenever possible. We also adopted SVN (see http://subversion.apache.org/) 
as a source code manager.
The software is developed for the UNIX environment. Another early decision 
was to have a pipeline working without a database (DB).
The input data, and intermediate and final products, are distributed in 
an intuitive directory tree, so that, even without the pipeline GUI, it is easy to mine
the filesystem for some particular result. 

The present tree structure was reached after a long improving process. 
We decided to avoid the DBs to make our pipeline as simple as possible
(also from an administrative point of view) and flexible enough to
allow software modifications and changes of the data processing 
model while remaining consistent with the previous SW versions.

\section{Science verification}
VST--Tube was developed as multi-instrument pipeline and tested on
a wide range of science cases. In this Section we present some
examples, either published or still in preparation, pointing out the
peculiarities of each case and the quality of the final results.

\subsection{CFHT12K F14 field}
We used VST--Tube to reduce a very difficult set of early 4
$deg^2$ CFHT12k images (VVDS-14h field \citep{garilli},
\citep{mccracken}): very sparse (distributed over 51 nights), with
few calibration data, with two CCDs (no. 0,6) of the mosaic which had
to be excluded (the pipeline was modified to handle mosaics with
unequal number of CCDs), with small offsets in the dithering for
both the science images and the calibration data (making it difficult to
remove objects), with almost no overlap in neighboring pointings,
very strong patterns in the raw images, many non--photometric
nights, and many corrupted headers or incomplete information.

\begin{figure}[h]
  \includegraphics[width=0.4\textwidth]{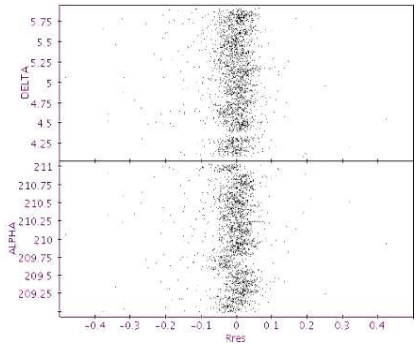}
  \caption{VST--Tube photometry of CFHT12k F14 field in R filter against SDSS's as a function of RA and Dec.
  Residual differences are better than $5\%$.}
  \label{f14}
  \end{figure}

In spite of the difficulties, the images where successfully processed 
and calibrated, and the results published  \citep{lamareille}. 
To verify the photometric accuracy, we made a comparison with 
SDSS observations of the same field. Figure \ref{f14} shows 
the magnitude residuals against SSDS's as function of Right 
Ascension and Declination. The standard deviation is better than 0.05 mag in the R band.

\subsection{WFI@2.2m NGC3115}
A very tight test on the accuracy of the instrumental signature
removal is through the brightness surface profiles of nearby
galaxies. A good example is that of the galaxy NGC 3115 since,
with a standard isophotal major diameter of 8.63', it cover most
of the field of view of the wide field imager WFI@2.2m. WFI
archive data in BVR bands of NGC3115 were retrieved and processed with
VST--Tube. 
 \begin{figure}[h]
 \centering
  \includegraphics[width=0.4\textwidth]{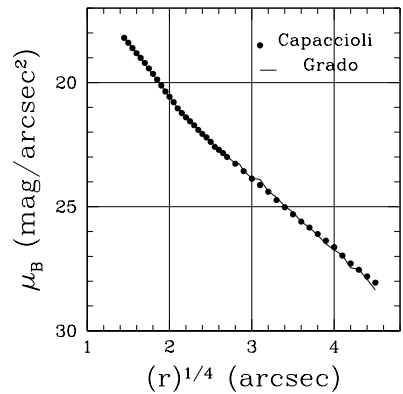}
  \caption{Surface brightness profile comparison of NGC3115. 
  The solid line is the profile measured on WFI@2.2m data in B 
  Johnson band (4200 seconds total exposure time). The dotted profile  
  comes from  Capaccioli et al. (see text)}
  \label{ngc3115}
  \end{figure}
A critical aspect was the harmonization
of the gain correction among the CCDs of the WFI mosaic. The
difficulties  arise from the lack of suitable data to produce a
superflat (in order to better flatten the instrumental background) and
from the size of the galaxy, which covers most of the field of
view. In order to test the quality of the VST--Tube results, the
surface brightness profiles along the main axis of the galaxy were
extracted and compared with a very accurate and deep photometric study by Capaccioli
\citep{capaccioli}, where wide--field photographic plates and CCD
data for more than 35000 seconds of observations were used. The
figure \ref{ngc3115}  \citep{capaccioli3115} shows the comparison of the light profiles
along the minor axis measured on a mosaic produced with VST--Tube
(solid line) (total exposure time 4200 seconds) and literature profile \citep{capaccioli}. The profiles agreed within .1
magnitude down to a surface brightness of 27.5 $mag
\ {\rm arcsec}^{-2}$.

\subsection{WFI@2.2m NGC6723}

\begin{figure}[h]
  \includegraphics[width=0.4\textwidth]{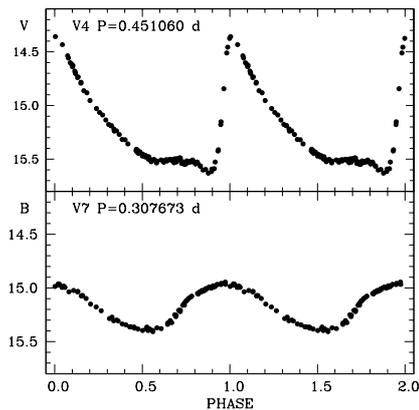}
  \caption{VST--Tube processing of the globular cluster NGC6723. Light curves of two RRLyrae in the field.}
  \label{6723}
  \end{figure}

The last example is the processing of a crowded stellar field: the globular 
cluster (GC) NGC6723 (Ripepi et al. in preparation). WFI BVI images with 120 phase points in each band 
were processed and the catalogs automatically extracted with PSF photometry.  The 
final mosaics, one for each exposure,  were registered also in pixels in order to have 
catalogs with sources in the same physical position and make easier the subsequent 
analysis with Stetson's tools \citep{stetson}. Figure \ref{6723} shows, as an example, two 
RRLyrae light curves found in the cluster. The rms of the residuals with respect to truncated Fourier series 
fits is better than $1\%$.  

\section{Acknowledgement}
The seed of this work was planted several years ago at the European Southern Observatory \citep{gradobeowfi}. 
For that reason one of the author (AG) wish to thanks Peter J. 
Quinn, Miguel Albrecht, and Andreas Wicenec. We also acknowledge the early contribution by  
Alfredo Volpicelli and Massimiliano Desiderio.  
We thank the Terapix group for providing QualityFits, 
Simone Zaggia for useful discussions and for providing the  
PSF photometry pipeline. Thanks to Mario Radovich for 
providing Photcal (the tool for the absolute photometric calibration) and to Mike Pavlov for the useful suggestions and discussions.
This work has been partially funded by the Regione Campania and by the Banco Napoli -- Fondazione, for which we thank the President, prof. A. Giannola.

\bibliographystyle{aa}

\end{document}